\documentclass[11pt]{article}
\usepackage{jheppub} 
\usepackage{mathtools}
\usepackage{physics}
\usepackage{amsmath}
\usepackage{amsfonts}
\DeclareMathOperator{\Ai}{Ai}

\makeatletter
\gdef\@fpheader{}
\makeatother

\begin{document}


\title{Analysis of Quartic Potential Problems Using Fredholm Integral Operators with Airy Functions}

\author{Ori J. Ganor$^1$}

\affiliation{$^1$ Leinweber Institute for Theoretical Physics, Department of Physics,
University of California, Berkeley, CA 94720, USA}

\emailAdd{ganor@berkeley.edu}

\vspace{10pt}

\abstract{
Fredholm integral operators that commute with the Hamiltonians of certain quantum mechanical systems with quartic potentials are introduced in closed analytical form. The operators are expressed in terms of the Airy function, and their eigenvalues fall off exponentially. Using these operators we obtain an exact dual description in terms of infinite one-dimensional chains. The systems discussed include the anharmonic quartic oscillator as well as multivariable potentials and higher dimensional systems. We compute a few moments of the Fredholm operators, using matrix models in the higher dimensional case, and we show how those can be used to approximate the energy spectrum and obtain estimates of the asymptotic behavior of the wavefunctions.
}

\vspace{2pc}


\maketitle


\def\be{\begin{equation}}
\def\ee{\end{equation}}
\def\bear{\begin{eqnarray}}
\def\eear{\end{eqnarray}}
\def\nn{\nonumber}
\newcommand{\secref}[1]{\S\ref{#1}}
\newcommand{\figref}[1]{Fig.~\ref{#1}}
\newcommand{\appref}[1]{Appendix~\ref{#1}}
\newcommand{\tabref}[1]{Table~\ref{#1}}

\newcommand{\R}{\mathbb{R}}

\newcommand\rep[1]{{\mathbf{#1}}} 
\newcommand\brep[1]{{\overline{\mathbf{#1}}}} 

\def\OpO{{\hat{O}}} 
\def\gQFT{{\mathbf{g}}}


\def\al{{\alpha}}
\def\gl{{\lambda}}
\def\aA{{\mathfrak{a}}}
\def\bA{{\mathfrak{b}}}
\def\hA{{\mathfrak{h}}}

\def\vz{{\mathfrak{z}}}
\def\vw{{\mathfrak{w}}}
\def\ChainN{{\mathfrak{n}}} 
\def\ChainM{{\mathfrak{m}}} 
\def\Normalization{{\mathcal Z}}
\def\GO{{\mathfrak{G}}}

\def\OpK{\mathcal{K}}
\def\OpKp{\mathcal{K}_{+}}
\def\OpKm{\mathcal{K}_{-}}
\def\OpKPM{\mathcal{\widetilde{K}}}

\def\OpKpMat{{\mathfrak{K}_{+}}}
\def\OpKMatH{{\mathfrak{S}}}

\def\OpH{{\hat{H}}}
\def\Ox{{\hat{x}}}
\def\Op{{\hat{p}}}

\def\mI{{m}}
\def\mMax{{M}}
\def\Hermite{{H}}
\def\matK{{\mathrm{K}}}
\def\ml{{\mu}}

\def\Ncutoff{{N}}
\def\Hilb{{{\mathcal H}}}
\def\ACS{{\Phi}}
\def\AFlat{{\mathcal{A}}}
\def\ActionL{{\Lambda}}
\def\cDF{{{\mathcal D}}} 
\def\vecVirial{{\mathrm{V}}} 
\def\wz{{\mathfrak{w}}}
\def\ApproxE{{\widetilde{E}}}
\def\matH{{\Theta}} 

\def\BinvH{{\kappa}}
\def\aaV{{\mu}}
\def\AinvH{{\gamma}}

\def\OneLoopE{{\widetilde{\widetilde{E}}}}
\def\Period{{\mathfrak{p}}}
\def\TadPole{{\mathfrak{T}}}
\def\SelfT{{\Sigma}}
\def\ml{{\mathbf{m}}}
\def\hA{{\mathbf{h}}}

\def\AsympC{{\mathrm{C}}}
\def\matK{{\mathrm{K}}}
\def\KXi{{\Xi}}
\def\vecX{{\mathrm{X}}}
\def\vecY{{\mathrm{Y}}}
\def\matZ{{\mathrm{Z}}}
\def\matW{{\mathrm{W}}}

\def\KPhi{{\Phi}}

\def\ul{{\mu}}
\def\uZ{{\mathrm{u}}}
\def\uA{{\mathfrak{c}}}
\def\FKer{{\mathrm{F}}}
\def\nr{{r}}

\def\evK{{\kappa}}
\def\evHK{{\gamma}}
\def\evW{{w}}

\def\gAAA{{g}}

\def\mr{{s}}

\def\xcp{{\mathrm{u}}}
\def\bxcp{{\overline{\xcp}}}
\def\vecU{{\mathrm{U}}}
\def\bvecU{{\overline{\vecU}}}
\def\vecV{{\mathrm{V}}}
\def\bvecV{{\overline{\vecV}}}
\def\matHK{{\mathrm{M}}}

\def\matW{{\mathrm{W}}}

\def\qChg{{q}}

\def\Id{{\mathbf{I}}} 

\def\Gz{{\mathrm{G}}}
\def\RGz{{\widetilde{\mathrm{G}}}}

\def\muEVHK{{\varpi}}



\section{Introduction and setup}

The purpose of this paper is to demonstrate that certain quantum mechanical problems with quartic potentials possess, in addition to standard parity and time reversal symmetries, another operator that commutes with the Hamiltonian. The operator is a Fredholm integral operator with a kernel function that can be expressed either in closed form or as an integral, and we will use it to derive analytical results for the energy spectrum. The systems that we aim to analyze are multivariable and interacting, with a potential that contains both quartic and quadratic terms, but these will be explored in the second part of the paper, in \secref{sec:Other}. The first part of the paper develops and demonstrates the techniques for the basic one-dimensional quantum mechanical problem with a quartic potential. 

Consider the one-dimensional quantum mechanical system with Hamiltonian
\be\label{eqn:H}
\OpH = -\frac{d^2}{dx^2}+\al x^2 + \frac{1}{2}\gl x^4,
\ee
where $\gl>0$ and $\al$ are given constants, and $x$ is in the range $-\infty<x<\infty$.
We define the operator $\OpKp$ by
\be\label{eqn:defOpKp}
\OpKp\psi(x) = \int_{-\infty}^\infty\Ai(\aA+\bA x^2+\bA y^2)\psi(y)dy,
\ee
where
\be\label{eqn:AiDef}
\Ai(u)=\frac{1}{2\pi}\int_{-\infty}^\infty e^{i u\vz +\frac{1}{3}i\vz^3}d\vz
\ee
is the Airy function, and 
\be\label{eqn:aAbA}
\aA=\gl^{-2/3}\al,
\qquad
\bA=\frac{1}{2}\gl^{1/3}.
\ee
$\OpKp$ commutes with $\OpH$, as is easily checked using the relation $\Ai''(u)=u\Ai(u)$.


The Hamiltonian \eqref{eqn:H} has been studied extensively, both analytically and numerically. The parameter $\bA$ can be absorbed by rescaling $x$, and the energy levels $E_k$ are proportional to $\bA$, with $E_k/\bA$ a function of $\aA$ only. The analytical behavior of $E_k/\bA$ as a function of (complex) $\aA$ was analyzed in the seminal works \cite{Bender:1969si,Simon:1970mc,Eremenko:2008}, where it was found that the parity even and parity odd energy spectra can be regarded as multivalued functions of $\aA$ that describe two separate Riemann surfaces with energy levels crossing at branch points of ramification index $2$. In \cite{Voros:1994}, exact algebraic identities satisfied by the spectral determinants for even and odd states [i.e., $\prod_{k=0}^\infty(1+E/E_{2k})$ and $\prod_{k=0}^\infty(1+E/E_{2k+1})$] for $\aA=0$ were discovered, using the controlled behavior of the determinants under $x\rightarrow e^{2\pi i/3}x$. This culminated with the discovery of the ODE/IM correspondence \cite{Dorey:1998pt,Dorey:2007zx}.

While standard perturbation theory does not converge for any $\gl$ \cite{Bender:1969si}, it can be Borel-resummed \cite{Bogomolny:1980ur,Zinn-Justin:1981qzi}, and so can the WKB expansion, leading to exact quantization conditions (``exact WKB'') by tracing the asymptotic behavior of the wavefunction from $x=+\infty$ to $x=-\infty$ through a path in the complex plane incorporating Stokes phenomena \cite{Voros:1983,Delabaere:1997srq}. (This was extended to potentials of odd degree \cite{Voros:1999rpg}, and the trans-series of perturbation theory was related to the exact WKB formulation in \cite{Bucciotti:2023trp}.) More recently, it was suggested \cite{Edery:2024chj,Edery:2025off} that truncating the Feynman path integral of a quartic potential with a cutoff range for $x$ also leads to convergent series. Given the energy $E$, the wavefunctions themselves can be expressed in terms of Heun functions, as found in \cite{Dong:2019wxa}.

Numerically, the variational method can achieve impressively high accuracy for energy eigenvalues and eigenfunctions \cite{Hioe:1978jj,Taseli:1996,Okun:2020,Turbiner:2021fgs}, with an accuracy of hundreds of digits for energy levels reported in some cases. Another successful approximation technique starts by converting Schr\"odinger's equation to a nonlinear Riccati equation that is then solved iteratively \cite{Liverts:2006}. In \cite{Turbiner:2020jqk,Turbiner:2021tmm} an approximate expression for the wavefunctions was reported, providing an accuracy of $\sim 10^{-10}$ for the first few energy levels and a large range of coupling constants. (More details on these approximation techniques can be found in \cite{Turbiner:2023book}.)
In \cite{Han:2020bkb} a ``bootstrap'' technique whereby a large number of inequalities on expectations values of operators in energy eigenstates are derived as a function of energy, which are then used to approximate the energy and expectation values themselves. Another approximation technique proposed in \cite{Caffarel:2023ppy,Caffarel:2024yry} begins by approximating the Feynman path integral with a suitably chosen Gaussian function, leading to an analytical expression with reported relative error of a few percents.

The spectrum of \eqref{eqn:H} has also been studied along nontrivial paths in the complex $x$-plane. For suitably chosen paths, $\gl<0$ is allowed when parity and time reversal symmetry guarantee a real spectrum \cite{Bender:1998ke,Bender:2007nj}. The $\gl<0$ system shows interesting thermodynamical behavior \cite{Romatschke:2020qfr} and the variational method has been adapted to that case in \cite{Romatschke:2026stg}.
The quartic potential with an additional linear term $n\sqrt{2\gl} x$ was studied in \cite{Bertola:2023}, for a natural number $n$, and along a path in the complex plane where $e^{x^3}$ decays at infinity. Solutions that can be expressed as polynomials times a simple decaying exponential factor were then found.

The approach taken in this paper is different from all those mentioned above.
As a Fredholm operator whose kernel decays exponentially fast, the eigenvalues of $\OpKp$ also fall off exponentially. As we will see, we can therefore use its moments to obtain good approximate expressions for the low-lying energy levels, as well as exact analytical results about the spectrum, which involve the exponentially decaying eigenvalues of $\OpKp$. 
We will, however, restrict to $\gl>0$ as the Airy kernel in \eqref{eqn:defOpKp} does not decay on at least one side of the path suggested in \cite{Bender:1998ke} for any choice of the cubic root $\bA$ in \eqref{eqn:aAbA}.

We begin in \secref{sec:EvsMoments} by analyzing the asymptotic behavior of the eigenvalues of $\OpKp$, and we compute its first three moments. Using $\OpKp$, we also present a new analytic expression for the constant prefactor in the asymptotic WKB approximation of an eigenfunction for the system with Hamiltonian \eqref{eqn:H}.
In \secref{sec:Chain} we calculate expectation values of operators $\OpO$ in the system by expanding expressions of the form $\tr(\OpO\OpKp^{\ChainN})$ using the integral representation \eqref{eqn:AiDef} of the Airy function. This leads to a ``dual'' description of the system whereby the ``fundamental'' variables are a chain of complex numbers $\vz_1,\dots,\vz_\ChainN$, and basic operators of the form $x^k p^m$ (where $p=-i d/dx$ and $k,m\ge 0$) are mapped to rational functions of the $\vz_i$'s. Since \eqref{eqn:defOpKp} is trivial on parity-odd states, we need a separate discussion for those, which we include in \secref{sec:ParityOdd}.
In \secref{sec:Other} we extend \eqref{eqn:defOpKp} to potentials of the form $\al x^2 + \frac{1}{2}\gl x^4 +\ml/x^2$ and to interacting systems with more than one variable, where the Airy function is replaced by a more complicated matrix integral. We then use matrix model techniques to calculate the first moment of the Fredholm operator. The multivariable systems can, in principle, be extended to quantum field theories with quartic interactions, but the interactions are nonlocal, as we discuss in \secref{sec:Disc}.

\subsection*{Note added}
After this paper was completed, I learned that the operator $\OpKp$ has been discovered before, and is a special case of a class of commuting operators studied in \cite{Halln_s_2015}, which also has some overlap with our observations on the asymptotic form of the wavefunction in \secref{subsec:AsympB}. I am indebted to Daniil Radamovich for bringing \cite{Halln_s_2015} to my attention.


\section{Eigenvalues and moments of $\OpKp$}
\label{sec:EvsMoments}

Let $\ket{n}$ ($n=0,1,2,\dots$) be the $n^{th}$ eigenstate of $\OpH$, with $n$ even (odd) corresponding to even (odd) wavefunctions. Since for any real $\aA$ all the eigenspaces of $\OpH$ are one-dimensional, and since $[\OpH,\OpKp]=0$, these $\ket{n}$'s are also eigenstates of $\OpKp$. For odd $n=2k+1$ we have $\OpKp\ket{2k+1}=0$, and for even $n=2k$ we define $\mu_{2k}$ to be the eigenvalue:
\be\label{eqn:mudef}
\OpKp\ket{2k}=\mu_{2k}\ket{2k}.
\ee
Following similar arguments to those presented in \cite{Simon:1970mc} for the analyticity of $E_k$ as a function of $\aA$, one finds that $\mu_{2k}$ are analytic in $\aA$ on the real axis. The argument is in fact much simpler for $\OpKp$, because it is a compact hermitian operator that depends holomorphically on $\aA$, and Rellich's theorem on the analyticity of the eigenvalues on the real axis for such a $1$-parameter family of compact operators applies. Extending to complex $\aA$, we have that $\mu_{2k}(\aA)$ are analytic except at the branch points where $E_{2k}(\aA)$ are ramified. This also follows immediately from the fact that the wavefunctions of $\OpH$ are analytic in $\aA$ except at the branch points \cite{Simon:1970mc}. Thus, like the $E_{2k}(\aA)$'s in \cite{Bender:1969si,Simon:1970mc,Eremenko:2008}, the $\mu_{2k}(\aA)$'s combine to a holomorphic function defined on the same ramified cover of the complex $\aA$-plane.

\subsection{Asymptotics}
\label{subsec:AsympB}

Using \eqref{eqn:mudef}, we can find the asymptotic behavior of the wavefunction for large $|x|$. Let $\psi_{2n}(x)$ be the $(2n)^{th}$ eigenfunction of the Hamiltonian. 
From the asymptotic form of Schr\"odinger's equation we easily find
$$
\psi_{2n}(x)\approx\frac{\AsympC_{2n}}{\bA|x|}e^{
-\frac{2}{3}\bA^{3/2}|x|^3-\bA^{1/2}\aA|x|}.
$$
The constant $\AsympC_{2n}$ can be determined by numerically integrating Schr\"odinger's equation, but using \eqref{eqn:mudef} we can find an analytical expression. Since $\psi_{2n}$ is also an eigenfunction of $\OpKp$, we have from \eqref{eqn:mudef},
\be\label{eqn:mu2npsi}
\mu_{2n}\psi_{2n}(x) = \int\Ai(\aA+\bA x^2+\bA y^2)\psi_{2n}(y)dy.
\ee
When $\bA x^2\gg 1$, the Airy function in the integrand is sharply peaked near $y=0$, and from its asymptotic expansion we find
$$
\psi_{2n}(x)\approx\frac{\psi_{2n}(0)}{2\bA\mu_{2n}|x|}e^{
-\frac{2}{3}\bA^{3/2}|x|^3-\bA^{1/2}\aA|x|},\qquad(\bA x^2\gg 1).
$$
For $n\gg 1$ we can compare this asymptotic expression to the WKB approximation and deduce the asymptotic fall-off of $\mu_{2n}$. For example, for $\aA=0$ we find 
\be\label{eqn:muWKB}
\mu_{2n}\approx\sqrt{\frac{2}{\pi\bA}} (-1)^{n} \left\lbrack \frac{\Gamma(\frac{1}{4})^2}{12(n+\frac{1}{4})} \right\rbrack^{1/3} e^{-(n+\frac{1}{4})\pi},
\qquad(\al=0,\,n\gg 1),
\ee
where we also used the WKB approximation for the energy
\be\label{eqn:EWKB}
E_{2n}\approx\left\lbrack\frac{12(n +\frac{1}{4})}{\Gamma(\frac{1}{4})^2} \right\rbrack^{4/3}\pi^2\bA,
\qquad(\al=0,\,n\gg 1).
\ee
Moreover, for any $n$ and $\aA$, we can find an expression for $\mu_{2n}$ by setting $x=0$ in \eqref{eqn:mu2npsi}, and we get the exact result for the asymptotic coefficient
\be\label{eqn:AsympC2n}
\AsympC_{2n}=\frac{\psi_{2n}(0)^2}{2\int\Ai(\aA+\bA x^2)\psi_{2n}(x)dx}\,.
\ee
Turbiner and del Valle \cite{Turbiner:2020jqk} developed a variational method ansatz that captures the $\sim\exp(-\frac{1}{3}\sqrt{\frac{\gl}{2}}|x|^3)$ asymptotic fall-off of the wavefunction, which then allows an excellent numerical determination of the constant $\AsympC_{2n}$. Using \eqref{eqn:AsympC2n}, however, we can propose an alternative method based on a simpler expansion of the exact wavefunctions in a basis of the standard harmonic oscillator eigenstates. Let $\Hilb^{+}_{2\Ncutoff}$ be the Hilbert space of linear combinations of the first $\Ncutoff+1$ even eigenstates of the harmonic oscillator with frequency set to $1$, i.e., of wavefunctions of the form $\psi(x)=P_{2\Ncutoff}(x)\exp(-\frac{1}{2}x^2)$, where $P_{2\Ncutoff}$ is an even polynomial of degree at most  $2\Ncutoff$. 
In this basis, the asymptotic behavior is wrong [$\sim \exp(-\frac{1}{2}x^2)$], but using a good approximation to the eigenfunction $\psi_{2n}(x)$, we can use \eqref{eqn:AsympC2n} to get a good approximation for $\AsympC_{2n}$. 

In \tabref{tab:muECPsi} we list the numerical values of $\mu_{2n}$ and $\AsympC_{2n}$ for $\gl=1$, and $2n=0,2,\dots,10$, obtained with $\Ncutoff=110$. The table also shows the fast fall off of $\mu_{2n}$ with $n$. In fact, instead of diagonalizing the Hamiltonian truncated to $\Hilb^{+}_{2\Ncutoff}$, one can diagonalize $\OpKp$ trunctated to the same Hilbert space. As far as the numerical effort, this only requires computing integrals of the form $\int_0^\infty\Ai(\bA r^2+\aA)e^{-r^2/2} r^j dr$ for $j=0,\dots,2\Ncutoff$, and it provides somewhat more accurate matrix elements between the exact ground states of the anharmonic oscillator and the ground states of the harmonic oscillator.

\begin{table}[h]

\begin{tabular}{|r|l|l|l|}
\hline
\multicolumn{1}{|c|}{$2n$} &
\multicolumn{1}{c|}{$E_{2n}$} &
\multicolumn{1}{c|}{$\mu_{2n}$} &
\multicolumn{1}{c|}{$|\AsympC_{2n}|$} \\
\hline
$0$ & 
 $0.84160994895089553$ & 
$+0.6484374347635222907935588464$ & 
$0.768321581751830567$ \\
$2$ & 
 $5.91759137495864209$ &
$-2.0569786488805969643915814491\times 10^{-2}$ &
$20.72816978957739975$ \\
$4$ & 
 $12.9070198646233429$ &
$+7.4169787962381625808216184647\times 10^{-4}$ &
$5.217962527128960753\times 10^2$ \\
$6$ & 
 $21.0556615320428476$ &
$-2.8520141922285371520997442097\times 10^{-5}$ &
$1.276801071432163867\times 10^4$ \\
$8$ & 
 $30.0995058620524286$ &
$+1.1305244565104984479870766462\times 10^{-6}$ &
$3.080678162781396859\times 10^5$ \\
$10$ & 
 $39.8884156438820210$ &
$-4.5618425292267635149635751818\times 10^{-8}$ &
$7.370987158380766684\times 10^6$ \\
\hline
\end{tabular}

\caption{
Listed are the (rounded) energy $E_{2n}$, the $\OpKp$ eigenvalue $\mu_{2n}$, and the coefficient of the asymptotic expansion $\AsympC_{2n}$ for the first $6$ even energy levels of the Hamiltonian \eqref{eqn:H} with $\al=0$ and $\gl=1$.
}
\label{tab:muECPsi}
\end{table}

\subsection{First three moments}
\label{subsec:Moments}

For small $\ChainN$ we can find an explicit expression for the moment $\tr(\OpKp^\ChainN)=\sum_{k=0}^\infty\mu_{2k}^\ChainN$. For the first moment we get
\be\label{eqn:muSum}
\sum_{k=0}^\infty\mu_{2k}=\int_{-\infty}^\infty\Ai(2\bA x^2+\aA)dx=
2^{1/6}\pi\bA^{-1/2}\left\lbrack\Ai\left(2^{-2/3}\aA\right)\right\rbrack^2.
\ee
We note that it is annihilated by the differential operator
$
2\frac{\partial^4}{\partial\aA^4}-2\aA\frac{\partial^2}{\partial\aA^2}-3\frac{\partial}{\partial\aA}.
$

For the second moment we get
\be\label{eqn:SecondMoment}
\sum_{k=0}^\infty\mu_{2k}^2 = 
\iint
\Ai(\bA x^2+\bA y^2+\aA)^2
dx dy=
\frac{2\pi}{\bA}\int\Ai(r^2+\aA)^2 r dr
=-\frac{\pi}{\bA}\aA\Ai(\aA)^2+\frac{\pi}{\bA}\Ai'(\aA)^2.
\ee

The third moment is a bit more complicated.
For this we need to evaluate
\be\label{eqn:faAdef}
f(\aA)\equiv 
\iiint
\Ai(\bA x^2+\bA y^2+\aA)
\Ai(\bA x^2+\bA z^2+\aA)
\Ai(\bA y^2+\bA z^2+\aA)
dx dy dz.
\ee
With integration by parts, it can be checked that $f(\aA)$ satisfies the $7^{th}$ order differential equation
\be\label{eqn:7thOrderODE}
0 = 27 \aA^2 f + 2(9\aA^3-37)f' -114 \aA f'' -38 \aA^2 f'''+55 f^{(4)}
+22 \aA f^{(5)} -2f^{(7)},
\ee
whose solutions can be expressed in terms of contour integrals involving generalized hypergeometric functions ${}_2F_2$. The exact linear combination of the seven linearly independent solutions to \eqref{eqn:7thOrderODE} that matches the integral \eqref{eqn:faAdef} can then be selected by matching the asymptotic behavior near $\aA=\infty$ [which needs to be $f(\aA)\approx e^{-2\aA^{3/2}}/16\sqrt{2}\aA^{3/2}$, using the asymptotic expansion of the Airy function]. The result is then
\bear
\sum_{n=0}^\infty\mu_{2n}^3 &=&
\iiint\Ai(\bA x_1^2+\bA x_2^2+\aA)\Ai(\bA x_2^2+\bA x_3^2+\aA)\Ai(\bA x_3^2+\bA x_1^2+\aA)dx_1 dx_2 dx_3
\nn\\ &=&
\frac{1}{\bA^3}\int_{-\infty}^\infty\Bigl\{
\frac{\Gamma(\frac{1}{6})^2}{2^{8/3}\sqrt{3}\pi\Gamma(\frac{2}{3})}
\Bigl\lbrack{}_2F_2\left(\tfrac{1}{3},\tfrac{1}{3};\tfrac{2}{3},\tfrac{5}{6};-\tfrac{8}{27}i u^3\right)
\Bigr\rbrack
-\sqrt{\tfrac{2}{\pi}}e^{-\frac{\pi i}{4}}\sqrt{u}
\Bigl\lbrack {}_2F_2\left(\tfrac{1}{2},\tfrac{1}{2};\tfrac{5}{6},\tfrac{7}{6};-\tfrac{8}{27}i u^3\right)
\Bigr\rbrack
\nn\\ &&
-\frac{3\cdot 2^{2/3}\Gamma(\frac{2}{3})}{\Gamma(\frac{1}{6})^2}i u
\Bigl\lbrack {}_2F_2\left(\tfrac{2}{3},\tfrac{2}{3};\tfrac{4}{3},\tfrac{7}{6};-\tfrac{8}{27}i u^3\right)
\Bigr\rbrack 
\Bigr\} e^{\frac{i}{3}u^3+i\aA u}du.
\label{eqn:muMom3}
\eear
More details are provided in \appref{app:DerivationOfThirdMoment}.

The expressions simplify considerably for $\aA=0$ (or $\al=0$). For that special value of $\aA$ we find
\be\label{eqn:momentsmu}
\sum_{n=0}^\infty\mu_{2n} = 
\frac{3^{1/6}}{\sqrt{2\bA\pi}}\Gamma(\tfrac{7}{6}),
\qquad
\sum_{n=0}^\infty\mu_{2n}^2 = 
\frac{3^{-2/3}\pi}{\bA \Gamma(\frac{1}{3})^2}\,,
\qquad(\al=0),
\ee
and the third moment simplifies to
\bear\lefteqn{
\sum_{n=0}^\infty\mu_{2n}^3
=
9\pi^{-3/2}\left(\tfrac{3}{2}\right)^{1/3}\Gamma\left(\tfrac{7}{6}\right)^3\left\lbrack{}_3F_2\left(\tfrac{1}{3},\tfrac{1}{3},\tfrac{1}{3};\tfrac{2}{3},\tfrac{5}{6};\frac{8}{9}\right)\right\rbrack
}\nn\\
&&
-2\left(\tfrac{2}{3}\right)^{1/2}\left\lbrack{}_3F_2\left(\tfrac{1}{2},\tfrac{1}{2},\tfrac{1}{2};\tfrac{5}{6},\tfrac{7}{6};\frac{8}{9}\right)\right\rbrack
+\frac{2^{2/3}\Gamma\left(-\tfrac{1}{3}\right)^2}{3^{5/6}\Gamma\left(\tfrac{1}{6}\right)^2}\left\lbrack{}_3F_2\left(\tfrac{2}{3},\tfrac{2}{3},\tfrac{2}{3};\tfrac{7}{6},\tfrac{4}{3};\frac{8}{9}\right)
\right\rbrack,
\qquad(\al=0).
\nn\\ &&
\label{eqn:Moment3}
\eear
Note that the first and third generalized hypergeometric functions ${}_3F_2$ on the RHS of \eqref{eqn:Moment3} can be expressed (using an identity of Claussen) in terms of Gauss hypergeometric functions as
\be\label{eqn:ClausenF}
{}_3F_2\left(\tfrac{1}{3},\tfrac{1}{3},\tfrac{1}{3};\tfrac{2}{3},\tfrac{5}{6};\frac{8}{9}\right)=
\left\lbrack
{}_2F_1\left(\tfrac{1}{3},\tfrac{1}{3};\tfrac{5}{6};\frac{1}{3}\right)
\right\rbrack^2,
\quad
{}_3F_2\left(\tfrac{2}{3},\tfrac{2}{3},\tfrac{2}{3};\tfrac{4}{3},\tfrac{7}{6};\frac{8}{9}\right)=
\left\lbrack
{}_2F_1\left(\tfrac{2}{3},\tfrac{2}{3};\tfrac{7}{6};\frac{1}{3}\right)
\right\rbrack^2.
\ee

We can also straightforwardly calculate $\tr(\OpH\OpKp^\ChainN)$ for $\ChainN=1,2$ to obtain
\be\label{eqn:Emu1Sum}
\sum_{k=0}^\infty E_{2k}\mu_{2k}=
2^{-5/6}\bA^{1/2}\pi\left\lbrack
2^{2/3}\aA\Ai'(2^{-2/3}\aA)^2
-3\cdot 2^{1/3}\Ai'(2^{-2/3}\aA)\Ai(2^{-2/3}\aA)
-\aA^2\Ai(2^{-2/3}\aA)^2
\right\rbrack,
\ee
and
\be\label{eqn:Emu2Sum}
\sum_{k=0}^\infty E_{2k}\mu_{2k}^2=
\frac{\pi}{30}\left\lbrack
(27+8\aA^3)\Ai(\aA)^2
-4\aA\Ai(\aA)\Ai'(\aA)
-8\aA^2\Ai'(\aA)^2
\right\rbrack.
\ee
Again, for $\aA=0$ these expressions simplify considerably:
\be\label{eqn:momentsmuEZ}
\sum_{n=0}^\infty E_{2n}\mu_{2n} = 
\sqrt{\frac{3\bA}{8}}\,,
\qquad
\sum_{n=0}^\infty E_{2n}\mu_{2n}^2 = 
\frac{3^{2/3}\pi}{10\Gamma\left(\frac{2}{3}\right)^2}\,,
\qquad(\al=0).
\ee

The moments obviously give upper bounds on the ground state energy. For example, for $\al=0$,
\be\label{eqn:LowerBound}
E_0<
\frac{\sum_{n=0}^\infty E_{2n}\mu_{2n}^2}{\sum_{n=0}^\infty\mu_{2n}^2}
=\frac{3^{4/3}\Gamma\left(\frac{1}{3}\right)^2}{10\Gamma\left(\frac{2}{3}\right)^2}\bA,\qquad
(\al=0).
\ee
As we have seen in \secref{subsec:AsympB}, the sequence $(\mu_{2n})$ falls off exponentially, and the RHS of \eqref{eqn:LowerBound} is actually only off by $0.6\%$ from the true $E_0$. Moreover, approximating $\mu_{2n}\approx 0$ for $n\ge 2$ in \eqref{eqn:momentsmu} and \eqref{eqn:momentsmuEZ} and solving for $\mu_0,\mu_2, E_0, E_2$ gives an expression for $E_0$ that is less than  $0.07\%$ off. 
Numerically, we can further improve the accuracy by using the WKB approximation for the tail of the sums in \eqref{eqn:momentsmu} and \eqref{eqn:momentsmuEZ}. Thus, using \eqref{eqn:muWKB} and \eqref{eqn:EWKB}, instead of simply setting $\mu_{2n}=0$ for $n\ge 2$, gives an approximation for $E_0$ that is less than $0.002\%$ off (for $\al=0$). Furthermore, if we use \eqref{eqn:muWKB} to approximate $\mu_{2n}$ only for $n\ge 3$, then solve \eqref{eqn:momentsmu}-\eqref{eqn:Moment3} for $\mu_0,\mu_2,\mu_4$, and then solve \eqref{eqn:momentsmuEZ} with the approximations \eqref{eqn:EWKB} for $E_4, E_6, \dots$ and \eqref{eqn:muWKB} for $\mu_6, \mu_8,\dots$, we find an approximation for $E_0$ that is only about $10^{-6}$ off. Thus, using the moments is an efficient tool to get very good approximations for the ground state energy.

If we wish to extend formulas \eqref{eqn:Moment3} and compute $\sum E_{2n}\mu_{2n}^3$ we will need the integral
\be\label{eqn:gaAdef}
\gAAA(\aA)\equiv 
\iiint
x^4\Ai(\bA x^2+\bA y^2+\aA)
\Ai(\bA x^2+\bA z^2+\aA)
\Ai(\bA y^2+\bA z^2+\aA)
dx dy dz,
\ee
which we can then use in the virial theorem [see \eqref{eqn:VirialEV} below].
Similarly to \eqref{eqn:7thOrderODE}, it can be checked that $g(\aA)$ satisfies the linear ordinary differential equation
\be\label{eqn:g7thOrderODE}
0 = (220-9\aA^3)\gAAA + 2\aA(9\aA^3-52)\gAAA' -6\aA^2\gAAA'' -(38\aA^3+2)\gAAA'''-21\aA\gAAA^{(4)}+22\aA^2\gAAA^{(5)}+4\gAAA^{(6)} -2\aA\gAAA^{(7)},
\ee
but I do not know a closed form for the solution that corresponds to the integral \eqref{eqn:gaAdef}.


\section{A dual chain}
\label{sec:Chain}

We can use the compactness of $\OpKp$ to construct an approximate projection operator on the ground state from a large power $\OpKp^\ChainN$, as $\ChainN\rightarrow\infty$. But first, we need to establish that $|\mu_0|$ is the largest in absolute value.

Let $a_0\approx -2.33811$ be the greatest zero of the Airy function. Then, for $\aA>a_0$ the kernel $\Ai(\aA+\bA x^2+\bA y^2)$ is positive, the greatest eigenvalue of $\OpKp$ is positive and also largest in absolute value, and the corresponding wavefunction $\psi(x)$ can be taken as positive everywhere. This is because the corresponding state $\ket{\psi}$ can be taken as the normalized state (up to a constant phase) that maximizes the expectation value $\bra{\psi}\OpKp\ket{\psi}$. Since the wavefunction is positive, it must correspond to the ground state $\ket{0}$. Thus, for $\aA>a_0$ we find $\mu_0>|\mu_{2k}|$ for all $k\ge 1$. Empirically, it turns out that $\mu_0>|\mu_{2k}|$ (for all $k\ge 1$) continues to hold for $\aA>a_1\approx -3.26$ but at $\aA=a_1$ we have $\mu_0=-\mu_2$, and for $\aA<a_1$ we cannot assume that $|\mu_0|$ is the largest. To simplify matters, we will therefore assume that $\aA>a_1$ from now on.

Let $\OpO$ be an arbitrary operator, with matrix elements $\OpO_{mn}=\bra{m}\OpO\ket{n}$ in the energy basis, and matrix elements $\OpO(x,y)$ in the position basis. If $\aA>a_1$, we can then compute the expectation value of $\OpO$ in the ground state by computing $\tr(\OpO\OpKp^\ChainN)/\tr(\OpKp^\ChainN)$ for a large integer $\ChainN\rightarrow\infty$, since $\OpKp^\ChainN/\tr(\OpKp^\ChainN)$ approaches a projection operator on the ground state exponentially fast. We will now evaluate $\tr(\OpO\OpKp^\ChainN)$ and $\tr(\OpKp^\ChainN)$ using the explicit expression \eqref{eqn:defOpKp}, leading to an integral over $\ChainN$ complex variables depicted in \figref{fig:Chain} below.


\subsection{Construction}
\label{subsec:Factors}

We start with the integral representation of the trace,
\bear
\lefteqn{
\sum_{k=0}^\infty\OpO_{2k,2k}\mu_{2k}^\ChainN =
\tr(\OpO\OpKp^\ChainN) 
}\nn\\ &=&
\int
\OpO(x_\ChainN,x_1)
\Ai(\aA+\bA x_1^2+\bA x_2^2)
 \Ai(\aA+\bA x_2^2+\bA x_3^2)
 \cdots
  \Ai(\aA+\bA x_\ChainN^2+\bA x_1^2)dx_1\cdots dx_\ChainN.
  \nn\\ &&
  \label{eqn:IntAiChain}
\eear
Substituting \eqref{eqn:AiDef} in \eqref{eqn:IntAiChain} and integrating over $x_1,\dots,x_\ChainN$, we arrive at
\be\label{eqn:AiryChainP}
\sum_{k=0}^\infty\OpO_{2k,2k}\mu_{2k}^\ChainN
=
\frac{e^{\frac{i\ChainN\pi}{4}}}{(4\bA\pi)^{\ChainN/2}}
\int
\GO_\OpO(\vz_\ChainN,\vz_1)
\prod_{k=1}^\ChainN\frac{e^{\frac{i}{3}\vz_k^3+i\aA\vz_k}d\vz_k}{(\vz_k+\vz_{k-1})^{\frac{1}{2}}},
\ee
where, up to a constant, $\GO_\OpO(\vz,\vz')$ is the matrix element of $\OpO$ between states with wavefunctions $\exp(i\bA\vz x^2)$ and $\exp(-i\bA\vz' x^2)$, and we defined $\vz_0=\vz_\ChainN$. The integration path for $\vz_1,\dots,\vz_\ChainN$ can be defined parallel to the real axis, giving a small positive imaginary part to the variables, to avoid the singularity of the square root. 
Since $\mu_{2k}$ falls off exponentially fast with $k$, as we have seen in \secref{sec:EvsMoments}, all the matrix elements $\OpO_{2k,2k}$ can be recovered from the asymptotic expansion of the RHS  for $\ChainN\gg1$.

The RHS of \eqref{eqn:AiryChainP} can be visualized (see \figref{fig:Chain}) as a chain of $\ChainN$ nodes on which the factors $\exp(\frac{i}{3}\vz_k^3+i\aA\vz_k)$ reside, and links connecting consecutive nodes carry the factors $(\vz_k+\vz_{k-1})^{-\frac{1}{2}}$, with the first and last node connected by the $\OpO$-dependent factor $\GO_\OpO(\vz_\ChainN,\vz_1)$. The latter is  a simple rational function for simple operators constructed polynomially from $x$ and $p =-i d/dx$. Nondiagonal matrix elements of $\OpO$ can be similarly recovered by starting with $\tr(\OpO\OpKp^\ChainN\OpO\OpKp^\ChainM)$, and the {\it chain} can thus be considered dual to the original quartic Hamiltonian.

We note that the {\it chain} undergoes a phase transition at $\aA=a_1\approx -3.26$. For $\aA<a_1$ the ground state contribution to the LHS of \eqref{eqn:AiryChainP} is no longer dominant, since $\mu_0$ is no longer the largest in absolute value. For $\aA$ just below $a_1$, for example, the excited state $\ket{2}$ contributes the most, and the $\ChainN\rightarrow\infty$ limit calculates expectation values of operators in that state.

For future reference, we define the ``action'' in \eqref{eqn:AiryChainP} by
\be\label{eqn:ACS}
\ACS=\sum_{k=1}^\ChainN\left(\frac{1}{3}\vz_k^3+\aA\vz_k\right)
+\frac{i}{2}\sum_{k=1}^\ChainN\log\left(\vz_k+\vz_{k+1}\right).
\ee
A steepest-descent approximation of \eqref{eqn:AiryChainP}, as we shall see below,
where we expand around saddle points of $\ACS$ where all $\vz_k$ are in the upper half plane, provides a reasonably good approximation to the ground state energy even when $\al=0$.


To each parity-even operator constructed out of $x$ and $p=-id/dx$ we match an analytic function $\GO_\OpO(\vz_0,\vz_1)$. For example, for $\beta>0$,
\bear
e^{-\beta x^2}&\longrightarrow&
2e^{-\frac{i\pi}{4}}\sqrt{\bA\pi}
(\vz_0+\vz_1)^{1/2}\int_{-\infty}^\infty e^{i\bA\vz_0 x^2-\beta x^2+i\bA\vz_1 x^2}dx
=\left(
\frac{\vz_0+\vz_1}{\vz_0+\vz_1+\frac{i\beta}{\bA}}
\right)^{1/2},
\nn
\eear
from which we find by expansion
\be\label{eqn:x2k}
x^{2k}\longrightarrow\frac{i^k(2k)!}{2^{2k}k!\bA^k(\vz_0+\vz_1)^k}\,.
\ee
Similarly, for even powers of momentum we find
\be\label{eqn:p2k}
p^{2k}\longrightarrow
\frac{(2k)!}{k!}\left(-\frac{i\bA\vz_0\vz_1}{\vz_0+\vz_1}\right)^k,
\ee
and more generally,
\be\label{eqn:x2kp2m}
x^{2k}p^{2m}
\longrightarrow
(-i)^{m-k}
\frac{(2k)!(2m)!\bA^{m-k}}{(\vz_0+\vz_1)^{k+m}}
\sum_{l=0}^{\min(k,m)}\frac{(-1)^{l}\vz_0^{m-l}\vz_1^{m+l}}{4^{k-l}(k-l)! (m-l)! (2l)!},
\ee
and
\be\label{eqn:x2k1p2m1}
x^{2k+1}p^{2m+1}
\longrightarrow
(-i)^{m-k-1}
\frac{(2k+1)!(2m+1)!\bA^{m-k}}{(\vz_0+\vz_1)^{k+m+1}}
\sum_{l=0}^{\min(k,m)}\frac{(-1)^{l}\vz_0^{m-l}\vz_1^{m+l+1}}{4^{k-l}(k-l)! (m-l)! (2l+1)!},
\ee
can easily be obtained by expanding
$$
e^{i\alpha x}e^{i\beta p}
\longrightarrow
\exp\left(
\frac{i\bA\beta^2\vz_0\vz_1}{\vz_0+\vz_1}
-\frac{i\alpha\beta\vz_1}{\vz_0+\vz_1}
-\frac{i\alpha^2}{4\bA(\vz_0+\vz_1)}
\right).
$$
The Hamiltonian \eqref{eqn:H} corresponds to
\be\label{eqn:Hmapped}
\OpH\longrightarrow
\GO_\OpH(\vz_0,\vz_1)=
\frac{2i\bA(\aA-\vz_0\vz_1)}{\vz_0+\vz_1}
-\frac{3\bA}{(\vz_0+\vz_1)^2}.
\ee
Thus, the spectrum of the system can be computed from the $\ChainN\rightarrow\infty$ behavior of the chain depicted in \figref{fig:Chain}.
The ground state energy $E_0$ is obtained by matching the value associated with the chain to $E_0\mu_0^{\ChainN}$, as a function of $\ChainN$. The next level, $E_2$ is obtained by improving the approximation to $E_0\mu_0^{\ChainN}+E_2\mu_2^{\ChainN}$, etc.

\begin{figure}[h]
\begin{picture}(500,120)
\put(5,50){\begin{picture}(0,0)

\multiput(60,0)(100,0){4}{\circle*{6}}
\multiput(63,0)(100,0){3}{\line(1,0){94}}
\put(57,0){\line(-1,0){20}}
\multiput(32,0)(-8,0){3}{\circle*{2}}
\put(363,0){\line(1,0){20}}
\multiput(388,0)(8,0){3}{\circle*{2}}

\put(58,-42){$e^{\frac{i}{3}\vz_{\ChainN-1}^3+i\aA\vz_{\ChainN-1}}$}
\put(60,-35){\vector(0,1){30}}
\put(158,-42){$e^{\frac{i}{3}\vz_{\ChainN}^3+i\aA\vz_{\ChainN}}$}
\put(160,-35){\vector(0,1){30}}
\put(258,-42){$e^{\frac{i}{3}\vz_{1}^3+i\aA\vz_{1}}$}
\put(260,-35){\vector(0,1){30}}
\put(358,-42){$e^{\frac{i}{3}\vz_{2}^3+i\aA\vz_{2}}$}
\put(360,-35){\vector(0,1){30}}

\put(90,10){$\frac{1}{(\vz_{\ChainN-1}+\vz_{\ChainN})^{1/2}}$}
\put(190,10){$\frac{1}{(\vz_{\ChainN}+\vz_{1})^{1/2}}$}
\put(290,10){$\frac{1}{(\vz_{1}+\vz_{2})^{1/2}}$}

\thicklines
\put(163,0){\line(1,0){94}}
\put(170,50){$\frac{2i\bA(\aA-\vz_0\vz_1)}{\vz_\ChainN+\vz_1}
-\frac{3\bA}{(\vz_\ChainN+\vz_1)^2}$}
\put(170,48){\vector(0,-1){42}}

\end{picture}}
\end{picture}
\caption{
A section of the chain describing the system with Hamiltonian $\OpH=-\frac{d^2}{dx^2}+4\bA^2\aA x^2+4\bA^3 x^4$. The chain is a loop of $\ChainN\rightarrow\infty$ nodes.
To compute the spectrum we assign complex variables $\vz_1,\dots,\vz_\ChainN$ to the nodes. Node $k$ carries a factor $e^{\frac{i}{3}\vz_{k}^3+i\aA\vz_{k}}$, and the link between node $k$ and node $(k+1)$ caries a factor $(\vz_k+\vz_{k+1})^{-1/2}$. The link between, say, node $\ChainN$ and node $1$ (where the chain closes to a loop) carries an additional factor for the Hamiltonian, according to \eqref{eqn:Hmapped}, which is symbolized by the thick link. The product of the factors is then integrated on a path that runs slightly above the real axis, for each $\vz_k$.
The chain undergoes a phase transition at $\aA\approx -3.26$.
}
\label{fig:Chain}
\end{figure}

\subsection{The virial theorem}
\label{subsec:Virial}

It is instructive to check how the virial theorem is realized in the dual chain. By the virial theorem
\be\label{eqn:VirialEV}
\bra{n}\left(p^2-\al x^2-\gl x^4\right)\ket{n} = 0,\qquad n=0,1,2,\dots,
\ee
and since, by \eqref{eqn:x2k} and \eqref{eqn:p2k},
$$
p^2-\al x^2-\gl x^4\longrightarrow
-\frac{2i\bA\vz_0\vz_1}{\vz_0+\vz_1}
-\frac{2i\bA\aA}{\vz_0+\vz_1}
+\frac{6\bA}{(\vz_0+\vz_1)^2},
$$
the virial theorem follows from the identity
\be\label{eqn:DualVirial}
\int
\left\lbrack
\frac{i\vz_0\vz_1}{\vz_0+\vz_1}
+\frac{i\aA}{\vz_0+\vz_1}
-\frac{3}{(\vz_0+\vz_1)^2}
\right\rbrack
e^{i\ACS(\vz_0,\dots,\vz_{\ChainN-1})}\prod_{k=0}^{\ChainN-1}d\vz_k
=0,
\ee
which in turn follows from the fact that the integrand can be written as a total derivative.
To see this, define the flat connection (on $\mathbb{C}^{\ChainN}$)
\bear
\AFlat &=& i d\ACS = 
i\sum_{k=0}^{\ChainN-1}\vz_k^2d\vz_k
+i\aA\sum_{k=0}^{\ChainN-1}d\vz_k
-\frac{1}{2}\sum_{k=0}^{\ChainN-1}\frac{(2\vz_k+\vz_{k+1}+\vz_{k-1})d\vz_k}{(\vz_k+\vz_{k-1})(\vz_k+\vz_{k+1})}\,,
\nn
\eear
and the associated covariant derivative
$$
\cDF_k = \frac{\partial}{\partial\vz_k}+\AFlat_k.
$$
We also define the vector field
\bear
\lefteqn{
\vecVirial^k\frac{\partial}{\partial\vz_k} =
\sum_{k=1}^{\ChainN-2}
\frac{\vz_0(\vz_{k+1}-\vz_{k-1})}{(\vz_0+\vz_{n-1})(\vz_k+\vz_{k-1})(\vz_k+\vz_{k+1})}
\frac{\partial}{\partial\vz_k}
}
\nn\\
&&
+\left\lbrack
\frac{\vz_0-2\vz_{n-1}}{(\vz_0+\vz_{n-1})^2}
-\frac{\vz_0}{(\vz_0+\vz_{n-1})(\vz_0+\vz_1)}
\right\rbrack\frac{\partial}{\partial\vz_0}
+\left\lbrack
\frac{\vz_0}{(\vz_0+\vz_{n-1})(\vz_{n-1}+\vz_{n-2})}
-\frac{3\vz_0}{(\vz_0+\vz_{n-1})^2}
\right\rbrack\frac{\partial}{\partial\vz_{n-1}}\,.
\nn
\eear
Then, we calculate
\be\label{eqn:VirialDV}
\sum_{k=0}^{\ChainN-1}\cDF_k\vecVirial^k
=
\frac{6}{(\vz_0+\vz_{\ChainN-1})^2}-\frac{2i\aA}{\vz_0+\vz_{\ChainN-1}}-\frac{2i\vz_0\vz_{\ChainN-1}}{\vz_0+\vz_{\ChainN-1}}\,,
\ee
thus confirming \eqref{eqn:DualVirial}.


\subsection{Steepest descent approximation}
\label{subsec:SteepestDescent}

We can approximate the integral \eqref{eqn:AiryChainP} by expanding around the saddle points of $\ACS$ in the upper half of the complex plane. This approximation is valid, as we will argue below, in the same regime as perturbation theory ($\aA\gg1$), but gives surprisingly good results in the nonperturbative regime as well. The saddle point equation is
\be\label{eqn:SteepD}
0=\frac{\partial\ACS}{\partial\vz_k}
=\vz_k^2+\aA+\frac{i}{2}\left(\frac{1}{\vz_k+\vz_{k-1}}+\frac{1}{\vz_k+\vz_{k+1}}\right),
\ee
and the simplest solutions are those for which all $\vz_k$'s are equal. We set $\vz_k = i\wz$ and find
\be\label{eqn:wzCubic}
\wz^3-\aA\wz-\frac{1}{2}=0.
\ee
There is only one solution with positive $\wz$, and for that solution we can simply approximate
\be\label{eqn:OpO00}
\bra{0}\OpO\ket{0} = \lim_{\ChainN\rightarrow\infty}
\frac{\sum_{k=0}^\infty\bra{2k}\OpO\ket{2k}\mu_{2k}^\ChainN}{\sum_{k=0}^\infty\mu_{2k}^\ChainN}
\approx
\GO_\OpO(i\wz,i\wz).
\ee
We will now explore how good the approximation \eqref{eqn:OpO00} is for $\OpO=\OpH$. Using \eqref{eqn:Hmapped} with $\vz_k=i\wz$, we therefore propose an approximation for the ground state energy in the form
\be\label{eqn:ApproxE0}
\ApproxE_0=
\bA\left(\wz
+\frac{\aA}{\wz}
+\frac{3}{4\wz^2}
\right)
=
\bA\left(2\wz
+\frac{1}{4\wz^2}
\right),
\ee
where $\wz$ is the unique positive solution to \eqref{eqn:wzCubic}.
If we set $\aA=\al=0$ and $\gl=1$, for example, we get $\ApproxE_0=5\cdot 2^{-7/3}\approx 0.99$ while the exact solution is $E_0\approx 0.8416\cdots$ \cite{Okun:2020}.

For $\aA\gg 1$ we are in the perturbative regime, with $p^2+\al x^2$ as the unperturbed Hamiltonian and $\frac{1}{2}\gl x^4$ as the perturbation.
Then, it is not hard to check that $\ApproxE_0$ agrees with $0^{th}$ and $1^{st}$ order perturbation theory.
Indeed, in this limit we have
$$
\wz = \aA^{1/2}+\frac{1}{4\aA}-\frac{3}{32\aA^{5/2}}+O(\frac{1}{\aA^{4}}),
$$
and
\be\label{eqn:Pert3Terms}
\ApproxE_0 =
\bA\left(2\aA^{1/2}+\frac{3}{4\aA}-\frac{5}{16\aA^{5/2}}\right)+O(\frac{\bA}{\aA^{4}}),
\ee
and the first two terms of \eqref{eqn:Pert3Terms}, proportional to $2\aA^{1/2}+\frac{3}{4\aA}$, agree with the $0^{th}$ order ($E_0^{(0)}=2\bA\aA^{1/2}$) and $1^{st}$ order ($E_0^{(1)}=3\bA/4\aA$) corrections of perturbation theory, while the coefficient of the $3^{rd}$ term of \eqref{eqn:Pert3Terms} is different from that of $2^{nd}$ order perturbation theory.
Expanding the integral \eqref{eqn:AiryChainP} around the saddle point $\vz_k=i\wz$ will presumably yield better approximations.


\section{Parity odd states}
\label{sec:ParityOdd}

To study parity odd excited states, we need to take a different Fredholm operator, since only parity even states appear in \eqref{eqn:AiryChainP}. Luckily, there is a simple modification of \eqref{eqn:defOpKp} that is nontrivial on odd states and commutes with $\OpH$:
\be\label{eqn:defOpKm}
\OpKm\psi(x) = x\int y\Ai(\aA+\bA x^2+\bA y^2)\psi(y)dy.
\ee
We denote the eigenvalues of $\OpKm$ by $\mu_{2k+1}$, for $k=0,1,2,\dots$
The first two moments of $\OpKm$ and their values for the pure quartic potential ($\aA=0$) are
\bear
\lefteqn{
\sum_{k=0}^\infty\mu_{2k+1}=\int_{-\infty}^\infty x^2\Ai(2\bA x^2+\aA)dx
}\nn\\ &=&
\frac{\pi}{(2\bA)^{3/2}}\left\lbrack
2^{1/3}\Ai'(2^{-2/3}\aA)^2
-2^{-1/3}\aA\Ai(2^{-2/3}\aA)^2
\right\rbrack
\xrightarrow{\aA=0}
\frac{\pi}{2^{7/6}\cdot 3^{2/3}\Gamma(\frac{1}{3})^2\bA^{3/2}},
\label{eqn:OddMom1}
\eear
and
\bear
\lefteqn{
\sum_{k=0}^\infty\mu_{2k+1}^2 = 
\iint
x^2 y^2
\Ai(\bA x^2+\bA y^2+\aA)^2
dx dy
}\nn\\
\nn\\ &=&
\frac{\pi}{120\bA^3}\left\lbrack
(3-8\aA^3)\Ai(\aA)^2
+4\aA\Ai(\aA)\Ai'(\aA)
+8\aA^2\Ai'(\aA)^2
\right\rbrack
\xrightarrow{\aA\rightarrow 0}
\frac{\pi}{120\cdot3^{\frac{1}{3}}\Gamma(\frac{2}{3})^2\bA^3}\,.
\label{eqn:OddMom2}
\eear
Inserting the energy into the integrals we find
\bear
\lefteqn{
\sum_{k=0}^\infty E_{2k+1}\mu_{2k+1}
=
\int_{-\infty}^{\infty}\left(8\bA^2\aA x^2+12\bA^3 x^4\right)x^2\Ai(2\bA x^2+\aA)dx
}
\nn\\
&=&
2^{-5/6}\pi\bA^{-1/2}\left\{
\frac{1}{4}(9+2\aA^3)\Ai(2^{-2/3}\aA)^2
-2^{-2/3}\aA\Ai(2^{-2/3}\aA)\Ai'(2^{-2/3}\aA)
-2^{-1/3}\aA^2\Ai'(2^{-2/3}\aA)^2
\right\}
\nn\\
&&\xrightarrow{\aA=0}
\frac{3^{2/3}}{4\cdot 2^{5/6}\Gamma(\frac{2}{3})^2\bA^{1/2}}\,,
\label{eqn:OddEMom1}
\eear
and
\bear
\lefteqn{
\sum_{k=0}^\infty E_{2k+1}\mu_{2k+1}^2
=
\iint\left(8\bA^2\aA x^2+12\bA^3 x^4\right)x^2 y^2\Ai(\bA x^2+\bA y^2+\aA)^2dx dy
}\nn\\
&=&
\frac{\pi}{840\bA^2}\left\{
4\aA^2(8\aA^3+77)\Ai(\aA)^2
-(16\aA^3+175)\Ai(\aA)\Ai'(\aA)
-32\aA(\aA^3+10)\Ai'(\aA)^2
\right\}
\nn\\
&&\xrightarrow{\aA=0}
\frac{5\pi}{72\bA^2\Gamma(\frac{1}{3})\Gamma(\frac{2}{3})}\,,
\label{eqn:OddEMom2}
\eear
where we have used the virial theorem.
For the pure quartic oscillator, $\aA=0$, if we approximately solve \eqref{eqn:OddMom1},\eqref{eqn:OddMom2},\eqref{eqn:OddEMom1},\eqref{eqn:OddEMom2} for $\mu_1,\mu_3,E_1,E_3$, by setting $\mu_{2n+1}\approx 0$ for $n\ge 2$, we find a result for the first excited level $E_1$ that is only $0.1\%$ off the correct value.


As for the {\it chain} of \secref{sec:Chain}, the same steps that led to \eqref{eqn:AiryChainP} now give
\bear
\sum_{k=0}^\infty\OpO_{2k+1,2k+1}\mu_{2k+1}^\ChainN
&=&
\left(\frac{e^{\frac{3i\pi}{4}}}{4\bA\sqrt{\bA\pi}}\right)^{\ChainN}
\int
\GO^{(-)}_\OpO(\vz_\ChainN,\vz_1)
\prod_{k=1}^\ChainN\frac{e^{\frac{i}{3}\vz_k^3+i\aA\vz_k}d\vz_k}{(\vz_k+\vz_{k-1})^{\frac{3}{2}}},
\label{eqn:AiryChainM}
\eear
and so the dual chain for parity odd states is similar to that of \figref{fig:Chain}, but with the factors on links changed from $(\vz_k+\vz_{k-1})^{-\frac{1}{2}}$ to $(\vz_k+\vz_{k-1})^{-\frac{3}{2}}$, and the extra \eqref{eqn:Hmapped} factor associated with the Hamiltonian is now replaced with
\be\label{eqn:HmappedM}
\OpH\longrightarrow
\GO^{(-)}_\OpH(\vz_0,\vz_1)=
\frac{6i\bA(\aA-\vz_0\vz_1)}{\vz_0+\vz_1}
-\frac{15\bA}{(\vz_0+\vz_1)^2}.
\ee
To approximate the first excited state, the set of steps that led to \eqref{eqn:wzCubic} now give
\be\label{eqn:wzCubicM}
\wz^3-\aA\wz-\frac{3}{2}=0,
\ee
and instead of \eqref{eqn:ApproxE0} we get
\be\label{eqn:ApproxE1}
\ApproxE_1=
\bA\left(3\wz
+\frac{3\aA}{\wz}
+\frac{15}{4\wz^2}
\right)
=
\bA\left(6\wz
-\frac{3}{4\wz^2}
\right),
\ee
For $\al=0$ we thus find the approximate energy of the first excited state to be $\ApproxE_1=\frac{11}{4}(\frac{3}{2})^{1/3}\approx 3.148$ which is less than 5\% off from the exact result of $3.0158\ldots$ \cite{Okun:2020}.



%
%

\section{Other systems}
\label{sec:Other}

The technique can be extended to other systems as well.
In 1d, the Hamiltonian \eqref{eqn:H} can be generalized to
\be\label{eqn:Hm}
\OpH = -\frac{d^2}{dx^2}+\al x^2 + \frac{1}{2}\gl x^4 +\frac{\ml}{x^2},
\ee
where $x>0$ and $\ml\ge-\frac{1}{4}$. 
Hamiltonians of this type have been investigated, for example, in \cite{Bazhanov:1998wj,Dorey:1999uk} in connection with the ODE/IM correspondence \cite{Dorey:2007zx}.
To obtain a commuting Fredholm operator, we replace \eqref{eqn:defOpKm} with
\be\label{eqn:defOpKPM}
\OpKPM\psi(x) = \int_0^\infty x^\hA y^\hA\Ai(\aA+\bA x^2+\bA y^2)\psi(y)dy,
\ee
where $\hA = \frac{1}{2}+\sqrt{\ml+\frac{1}{4}}$.

In a different vein, we can also extend to higher dimensions.
Consider the Hamiltonian
\be\label{eqn:HmHighD}
\OpH = -\sum_{i=1}^\nr\frac{\partial^2}{\partial x_i^2}
+\sum_{i,j}\matK_{ij} x_i x_j+\frac{1}{2}\gl\left(\sum_{i=1}^\nr x_i^2\right)^2,
\ee
where $\matK$ is a symmetric matrix ($\matK^\intercal=\matK$), not necessarily positive definite, and $\gl>0$.
Clearly, only the eigenvalues of $\matK$ matter for the spectrum, but it will be convenient to keep it as a general symmetric matrix.
(The $O(\nr)$ symmetric Hamiltonian, with $\matK=0$, was used recently in \cite{Barberena:2025ibo}.)

Let $\vecX$ be an $\nr$-dimensional vector with components $x_1,\dots,x_\nr$, and let $\vecY$ be another  $\nr$-dimensional vector with components $y_1,\dots,y_\nr$. Let $\matZ$ be an arbitrary symmetric $\nr\times\nr$ matrix.
We define
\bear
\KXi(\vecX,\vecY,\matZ) &=&
\bA\left(\vecX^\intercal\matZ\vecX+\vecY^\intercal\matZ\vecY\right)
+\frac{1}{4\bA^2}\tr(\matZ\matK)
+\frac{1}{3}\tr(\matZ^3),
\nn
\eear
and let
$$
\KPhi(\vecX,\vecY)=\int e^{i\KXi(\vecX,\vecY,\matZ)}\prod_{i\le j}d\matZ_{ij}.
$$
Then, it is not hard to check with integration by parts that the following Fredholm operator, \be\label{eqn:OpKpMat}
\psi(\vecX)\mapsto
\OpKpMat\psi(\vecX)=\int\KPhi(\vecX,\vecY)\psi(\vecY)d^\nr\vecY,
\ee
commutes with the Hamiltonian \eqref{eqn:HmHighD}.
For $\nr=2$ we can actually integrate out all but one element of $\matZ$. For example, if we choose a basis where $\OpH$ takes the form
\be\label{eqn:Hn2}
\OpH = 
-\frac{\partial^2}{\partial x_1^2}
-\frac{\partial^2}{\partial x_2^2}
+\al(x_1^2+x_2^2)
+2\ul x_1 x_2
+\frac{1}{2}\gl(x_1^2+x_2^2)^2,
\ee
we find, after integrating out the component $\matZ_{12}$, changing variables to $\matZ_{11}\pm\matZ_{22}$, and integrating out $\matZ_{11}-\matZ_{22}$, that the following operator commutes with \eqref{eqn:Hn2}:
\be
\psi(x_1, x_2) \mapsto
\int\FKer(x_1, x_2; y_1, y_2)\psi(y_1, y_2)dy_1 dy_2,
\ee
where the Fredholm kernel can be expressed as an integral in one variable:
\bear
\lefteqn{
\FKer(x_1, x_2; y_1, y_2) =
\int_{-\infty}^\infty
\exp\Bigl\{
\frac{2i}{3}\uZ^3
+2i\aA\uZ 
+i\bA\uZ(x_1^2+x_2^2+y_1^2+y_2^2)
}
\nn\\ &&
-\frac{i}{8\uZ}\bA^2(x_1^2-x_2^2+y_1^2-y_2^2)^2
-\frac{i}{2\uZ}\left\lbrack
\bA(x_1 x_2+y_1 y_2)+\uA
\right\rbrack^2
\Bigr\}\frac{d\uZ}{\uZ},
\label{eqn:FK}
\eear
and where $\uA = \gl^{-2/3}\ul$, and $\aA$ and $\bA$ are given by \eqref{eqn:aAbA}.
We note that the integration path in \eqref{eqn:FK} needs to be deformed into the upper half complex plane in the vicinity of $\uZ=0$, in order to avoid the singularity. 

Returning to general $\nr$, we can repeat the same steps that led to \eqref{eqn:AiryChainP}, and arrive at a similar formula, but with $\vz_k$ replaced by an $\nr\times\nr$ matrix $\matZ_k$ and the factors $(\vz_k+\vz_{k-1})^{\frac{1}{2}}$ in the denominator replaced by $\sqrt{\det(\matZ_k+\matZ_{k-1})}$. Also, the normalization factor should be $(e^{i\pi/4}/2\sqrt{\bA\pi})^{\ChainN\nr}$, and $\aA\vz_k$ should be replaced with $\frac{1}{4\bA^2}\tr(\matZ_k\matK)$. 
For parity odd states, we can generalize \eqref{eqn:defOpKm} by replacing \eqref{eqn:OpKpMat} with
\be\label{eqn:OpKmMat}
\psi(\vecX)\mapsto\int\tr(\vecX^\intercal\vecY)\KPhi(\vecX,\vecY)\psi(\vecY)d^\nr\vecY.
\ee


Let us now investigate the first moment of $\OpKpMat$. I do not have a closed form formula, like \eqref{eqn:muSum}, for a generic (symmetric) coupling matrix $\matK$ in \eqref{eqn:HmHighD}, but if we restrict $\matK$ to have hermitian structure, then we can use basic techniques of matrix models \cite{Itzykson:1979fi} to reduce the integral that computes the first moment to an integral over the eigenvalues of $\matZ$, and then we can reach a closed form expression as follows.

First, by ``hermitian structure'' we mean that $\nr=2\mr$ and the $\nr$ variables $x_j$ ($1\le k\le\nr$) can be combined into $\mr$ complex variables $\xcp_k=x_k + i x_{k+\mr}$ ($1\le k\le\mr$), collected into an $\mr$-dimensional complex vector $\vecU$, so that the quadratic coupling in \eqref{eqn:HmHighD} can be expressed as
$$
\vecX^\intercal\matK\vecX = \vecU^\dagger\matHK\vecU,
$$
for some hermitian $\mr\times\mr$ matrix $\matHK$. The system then has an extra $U(1)$ symmetry acting as $\vecU\rightarrow e^{i\theta}\vecU$, and one can then check that there is an additional\footnote{I do not know if \eqref{eqn:OpKpMat} and \eqref{eqn:OpKMatH} are related in any way.} Fredholm operator that commutes with the Hamiltonian:
\bear
\lefteqn{
\psi(\vecU,\bvecU)\mapsto
}\nn\\
&&
\OpKMatH_0\psi(\vecU,\bvecU)=\int
\exp\left\lbrack
i\bA\left(\vecU^\dagger\matW\vecU+\vecV^\dagger\matW\vecV\right)
+\frac{i}{4\bA^2}\tr(\matW\matHK)
+\frac{i}{6}\tr(\matW^3)
\right\rbrack
\psi(\vecV,\bvecV)d\matW d\vecV d\bvecV,
\nn\\
&&
\label{eqn:OpKMatH}
\eear
where $\bvecU,\bvecV$ are the complex conjugate vectors of $\vecU,\vecV$, and the integral $\int(\cdots)d\matW$ is over hermitian matrices ($\matW=\matW^\dagger$).
In \eqref{eqn:OpKMatH} our convention is that if the elements of $\vecV$ are set to $y_k+i y_{k+\mr}$ ($1\le\mr$) then
$$
d\vecV d\bvecV = \prod_{j=1}^{2\mr} dy_j,\qquad
d\matW=\prod_{k}d\matW_{kk}\prod_{k<l}d\text{Re}\matW_{kl}\prod_{k<l}d\text{Im}\matW_{kl}.
$$
The operator $\OpKMatH_0$ is trivial on states with nonzero $U(1)$ charge, but on states with $\qChg>0$ units of charge we can use the following operator,
\bear
\lefteqn{
\psi(\vecU,\bvecU)\mapsto
\OpKMatH_\qChg\psi(\vecU,\bvecU)
}\nn\\
&=&\int
\left(\vecV^\dagger\vecU\right)^\qChg
\exp\left\lbrack
i\bA\left(\vecU^\dagger\matW\vecU+\vecV^\dagger\matW\vecV\right)
+\frac{i}{4\bA^2}\tr(\matW\matHK)
+\frac{i}{6}\tr(\matW^3)
\right\rbrack
\psi(\vecV,\bvecV)d\matW d\vecV d\bvecV.
\nn\\ &&
\label{eqn:OpKMatHQ}
\eear
This operator $\OpKMatH_\qChg$ also commutes with $\OpH$ and is only nontrivial on states with charge $\qChg$.

The first moment of $\OpKMatH_0$ can be calculated with a matrix integral
\be\label{eqn:trOpKMatH0}
\tr\OpKMatH_0 =
\left(\frac{i\pi}{2\bA}\right)^\mr
\int
\exp\left\lbrack
\frac{i}{6}\tr(\matW^3)
+\frac{i}{4\bA^2}\tr(\matW\matHK)
\right\rbrack
\frac{d\matW}{\det(\matW+i\epsilon\Id)}\,,
\ee
where $\epsilon$ is a small positive number, introduced to avoid the singularity $\det\matW=0$.
Let $\evHK_1,\dots,\evHK_\mr$ be the eigenvalues of $\matHK$. Then, using standard matrix model results \cite{Itzykson:1979fi} we can convert the integral to one over the eigenvalues $\evW_1,\dots,\evW_\mr$ of $\matW$:
\bear
\tr\OpKMatH_0 &=&
\pi^{\mr(\mr+1)/2}
i^{\mr(3-\mr)/2}
(2\bA)^{\mr(\mr-2)}
\int
\prod_{k<l}\left(\frac{\evW_k-\evW_l}{\evHK_k-\evHK_l}\right)
\exp\left\lbrack
\sum_{k=1}^\mr\left(\frac{i\evW_k^3}{6}+\frac{i\evHK_k\evW_k}{4\bA^2}\right)
\right\rbrack\prod_{k=1}^\mr\frac{d\evW_k}{\evW_k+i\epsilon}\,.
\nn\\ &&
\label{eqn:trOpKMatHInt}
\eear
For example, for $\mr=2$  we compute
$$
\tr\OpKMatH_0^{(\mr=2)} =
\frac{i\pi^3}{\evHK_1-\evHK_2}
\iint
\exp\left\lbrack
\frac{i}{6}(\evW_1^3+\evW_2^3)
+\frac{i}{4\bA^2}(\evHK_1\evW_1+\evHK_2\evW_2)
\right\rbrack
\frac{(\evW_1-\evW_2)d\evW_1 d\evW_2}{(\evW_1+i\epsilon)(\evW_2+i\epsilon)},
$$
and this integral can be expressed in closed form in terms of the generalized hypergeometric function ${}_1F_2$ as follows.
Define
\bear
\lefteqn{
\Gz_0(\zeta) =
-i\int_{-\infty}^\infty
\frac{e^{\frac{i}{6}\evW^3+i\zeta\evW}d\evW}{\evW+i\epsilon}
}\nn\\ &=&
2^{1/3}\cdot 3^{-1/6}\Gamma\left(\tfrac{1}{3}\right)
{}_1F_2\left(\tfrac{1}{3}; \tfrac{2}{3},\tfrac{4}{3};\tfrac{2}{9}\zeta^3\right)\zeta
-2^{-1/3}\cdot 3^{1/6}\Gamma\left(\tfrac{2}{3}\right)
{}_1F_2\left(\tfrac{2}{3}; \tfrac{4}{3},\tfrac{5}{3};\tfrac{2}{9}\zeta^3\right)\zeta^2
-\tfrac{2\pi}{3}.
\nn\\ &&
\label{eqn:defGz}
\eear
Then,
$$
\tr\OpKMatH_0^{(\mr=2)} =
\frac{2^{4/3}\pi^4}{\evHK_1-\evHK_2}\left\lbrack
\Ai\left(\frac{2^{1/3}\evHK_2}{4\bA^2}\right)\Gz_0\left(\frac{\evHK_1}{4\bA^2}\right)
-\Ai\left(\frac{2^{1/3}\evHK_1}{4\bA^2}\right)\Gz_0\left(\frac{\evHK_2}{4\bA^2}\right)
\right\rbrack.
$$
Returning to general $\mr$, we can replace $\prod_{k<l}(\evW_k-\evW_l)$ in \eqref{eqn:trOpKMatHInt} with a Vandermonde determinant, as is standard in matrix models (see, e.g., \cite{Anninos:2020ccj} for a recent review). Then, defining for $k\ge 1$,
\be\label{eqn:defGzn}
\Gz_k(\zeta) =
i^{k-1}\int_{-\infty}^\infty
\evW^{k-1} e^{\frac{i}{6}\evW^3+i\zeta\evW}d\evW
=2^{1+\frac{k}{3}}\Ai^{(k-1)}(2^{1/3}\zeta),
\ee
where $\Ai^{(k-1)}$ is the $(k-1)^{th}$ derivative of the Airy function, we can rewrite \eqref{eqn:trOpKMatHInt} as
\bear
\tr\OpKMatH_0 &=&
\frac{\pi^{\mr(\mr+1)/2}}{(2\bA)^\mr\prod_{k<l}(\evHK_l-\evHK_k)}
\det\left\lbrack
\Gz_{k-1}\left(\frac{\evHK_l}{4\bA^2}\right)
\right\rbrack_{1\le k,l\le\mr},
\label{eqn:trOpKMatVDM}
\eear
where the determinant is of an $\mr\times\mr$ matrix with elements $\Gz_{k-1}\left(\frac{\evHK_l}{4\bA^2}\right)$.

Inserting $\OpH$ into the trace \eqref{eqn:trOpKMatH0}, we find (using the virial theorem)
\bear
\lefteqn{\tr(\OpH\OpKMatH_0) =
\int\left\lbrack
2\vecU^\dagger\matHK\vecU+12\bA^3(\vecU^\dagger\vecU)^2\right\rbrack
\exp\left\lbrack
2i\bA\vecU^\dagger\matW\vecU
+\frac{i}{4\bA^2}\tr(\matW\matHK)
+\frac{i}{6}\tr(\matW^3)
\right\rbrack
d\matW d\vecU d\bvecU
}\nn\\ &=&
\left(\frac{i\pi}{2\bA}\right)^\mr
\int
\left\{
\frac{i}{2\bA}\tr(\matHK\matW^{-1})
-3\bA\left\lbrack\tr(\matW^{-1})\right\rbrack^2
-3\bA\tr(\matW^{-2})
\right\}
\frac{e^{\frac{i}{6}\tr(\matW^3)+\frac{i}{4\bA^2}\tr(\matW\matHK)}d\matW}{\det(\matW+i\epsilon\Id)}\,.
\nn\\ &&
\label{eqn:trOpKMatH0Calc}
\eear
This again can be simplified to an integral over the eigenvalues, 
\bear
\lefteqn{
\tr(\OpH\OpKMatH_0) = 
\pi^{\mr(\mr+1)/2}
i^{\mr(3-\mr)/2}
(2\bA)^{\mr(\mr-2)}
\int
\Bigl\{
\frac{i}{2\bA}
\left(1-\frac{\mr(\mr-1)}{2}+\frac{i}{4\bA^2}\sum_p\evHK_p\evW_p\right)\left(
\sum_{k}\frac{\evHK_k}{\evW_k}
\right)
}\nn\\ &&
+\bA\left(
\frac{\mr(\mr-1)}{2}
-\frac{i}{4\bA^2}\sum_k\evHK_k\evW_k
\right)
\left\lbrack
\sum_p\frac{1}{\evW_p^2}
-\left(\sum_p\frac{1}{\evW_p}\right)^2
\right\rbrack
-3\bA\left(\sum\frac{1}{\evW_k}\right)^2
-3\bA\sum\frac{1}{\evW_k^2}
\Bigr\}\times
\nn\\
&&
\prod_{k<l}\left(\frac{\evW_k-\evW_l}{\evHK_k-\evHK_l}\right)
\exp\left\lbrack
\sum_{k=1}^\mr\left(\frac{i\evW_k^3}{6}+\frac{i\evHK_k\evW_k}{4\bA^2}\right)
\right\rbrack\prod_{k=1}^\mr\frac{d\evW_k}{\evW_k}\,,
\nn
\eear
and for $\mr=2$ we find
\bear
\lefteqn{
\tr(\OpH\OpKMatH_0^{(\mr=2)}) =
\frac{i\pi^3}{\evHK_1-\evHK_2}
\int
\left\lbrack
\frac{i\evHK_1}{2\bA\evW_1}
+\frac{i\evHK_2}{2\bA\evW_2}
-\frac{6\bA}{\evW_1^2}
-\frac{6\bA}{\evW_2^2}
-\frac{4\bA}{\evW_1\evW_2}
\right\rbrack\times
}
\nn\\ &&
\exp\left\lbrack
\frac{i}{6}(\evW_1^3+\evW_2^3)
+\frac{i}{4\bA^2}(\evHK_1\evW_1+\evHK_2\evW_2)
\right\rbrack
(\evW_1-\evW_2)
\frac{d\evW_1 d\evW_2}{\evW_1\evW_2}.
\label{eqn:trHOpKMat2}
\eear
where the integration contours over $\evW_1$ and $\evW_2$ can be taken parallel to the real axis, in the upper half complex plane.
The explicit result of the integral \eqref{eqn:trHOpKMat2} can be found in \appref{app:trHOpKMat}.

The second moment of $\OpKMatH_0$ can also be reduced to an integral over the eigenvalues of a single hermitian $\mr\times\mr$ matrix as follows. First, we plug the integral expression \eqref{eqn:OpKMatH} into the definition of the second moment to obtain
\bear
\lefteqn{
\tr(\OpKMatH_0^2) =
\int
\exp\left\lbrack
i\bA\left(\vecU^\dagger\matW_1\vecU+\vecV^\dagger\matW_1\vecV\right)
+\frac{i}{4\bA^2}\tr(\matW_1\matHK)
+\frac{i}{6}\tr(\matW_1^3)
\right\rbrack
}\nn\\ &&
\exp\left\lbrack
i\bA\left(\vecV^\dagger\matW_2\vecV+\vecU^\dagger\matW_2\vecU\right)
+\frac{i}{4\bA^2}\tr(\matW_2\matHK)
+\frac{i}{6}\tr(\matW_2^3)
\right\rbrack
d\matW_1 d\matW_2 d\vecU d\bvecU d\vecV d\bvecV.
\nn
\eear
Then, we integrate over $\vecU,\vecV$ to get
\bear
\tr(\OpKMatH_0^2)&=&
\left(\frac{i\pi}{\bA}\right)^{2\mr}
\iint
\exp\left\{
\frac{i}{6}\tr(\matW_1^3+\matW_2^3)
+\frac{i}{4\bA^2}\tr\left\lbrack(\matW_1+\matW_2)\matHK\right\rbrack
\right\}\frac{d\matW_1 d\matW_2}{\det(\matW_1+\matW_2)^2},
\nn\\ &&
\eear
and after that we change variables to 
$$
\matW_{\pm}=\matW_1\pm\matW_2,
$$
using
$$
\tr(\matW_1^3+\matW_2^3) =
\tfrac{1}{4}\tr\left\lbrack
\matW_{+}^3+3\matW_{+}\matW_{-}^2\right\rbrack.
$$
We then find that the integral over $\matW_{-}$ is Gaussian, and the remaining integral over $\matW_{+}$ can be converted with standard matrix model techniques, as above, to an integral over the eigenvalues $\evW_1,\dots,\evW_\mr$ of $\matW_{+}$:
\bear
\tr(\OpKMatH_0^2)&=&
\frac{2^{\mr(\frac{3}{2}\mr-1)}
\pi^{\mr(\mr+\frac{3}{2})}
\bA^{\mr(\mr-3)}
e^{\frac{\pi i}{8}\mr(\mr+9)}
}{\prod_{k<l}(\evHK_k-\evHK_l)}
\int
e^{\frac{i}{24}\sum\evW_j^3+\frac{i}{4\bA^2}\sum\evW_j\evHK_j}
\prod_{k<l}\frac{\evW_k-\evW_l}{\evW_k+\evW_l}
\prod_{j=1}^\mr \frac{d\evW_j}{\evW_j^{5/2}}.
\nn\\ &&
\eear


\section{Discussion and outlook}
\label{sec:Disc}

As we have seen, a variety of quantum mechanical systems with quartic potentials (as well as some generalizations) possess nontrivial compact Fredholm integral operators that commute with the Hamiltonian. We have explored the properties of the spectrum of these compact operators and how they yield exact analytical results on the spectrum of the Hamiltonian. In particular, we found sum rules, such as \eqref{eqn:momentsmuEZ}, that can be solved approximately to give a good approximation for the ground state energy. For the pure one-dimensional quartic potential [\eqref{eqn:H} with $\al=0$], we found explicit expressions for the first, second, and third moments \eqref{eqn:momentsmu}-\eqref{eqn:Moment3}, and exact expressions for the first and second moments with a Hamiltonian insertion \eqref{eqn:momentsmuEZ}. It would be interesting to explore whether explicit expressions can be found for higher moments, or at least expressions that can be calculated efficiently numerically, and whether this can be used as a practical technique for numerical approximation of the spectrum, as suggested below \eqref{eqn:LowerBound}.
By evaluating large moments of the compact operator with other operator insertions, we found an exact dual description of the system as a chain where elements of the Weyl algebra of operators (constructed out of position and momentum) are replaced with rational functions of the complex variables of the chain [see \eqref{eqn:x2kp2m}-\eqref{eqn:x2k1p2m1}]. Then, identities such as the virial theorem can be recast in algebraic geometrical terms. The chain undergoes a phase transition at a value of the coupling constant $\aA=a_1$ that we found numerically as $\approx -0.326$. Below that value, the infinite chain no longer corresponds to the ground state of the anharmonic oscillator.

Do such commuting Fredholm integral operators also exist for quantum field theories with quartic potentials?
Our results naturally extend to certain interacting systems of quartic oscillators, with Hamiltonian \eqref{eqn:HmHighD}, as we have seen in \eqref{eqn:OpKpMat}. The continuum limit of a $(D+1)$-dimensional lattice model with quartic potential and Hamiltonian of the form \eqref{eqn:HmHighD} leads to a $(D+1)$-dimensional QFT with a nonlocal (quartic) interaction, with Lagrangian
$$
\tfrac{1}{2}\int\left(\partial_\mu\phi\partial^\mu\phi -m^2\phi^2\right)d^Dx dt
-\gQFT\int\left\lbrack\int\phi(t,x)^2d^Dx\right\rbrack^2 dt,
$$
where $\gQFT$ is a coupling constant, and $\phi$ is a scalar field.
This Lagrangian might be interesting to study in its own right, and indeed it has nontrivial radiative corrections \cite{GanorAndHe}, but it would be also interesting to look for a simple Fredholm operator in a local QFT. 
For example, do such operators exist for quantum mechanical systems with potentials of the form $V(x_1, x_2)=\frac{1}{2}\gl(x_1^4+x_2^4)+\mu x_1 x_2$, even to first order in $\mu$? If such explicit Fredholm operators can be found for local QFTs, they might also elucidate the remarkable proposal by Gaiotto and Witten of an operator that realizes S-duality on boundary conditions (or, equivalently, wavefunctionals) of $N=4$ Super Yang-Mills theory \cite{Gaiotto:2008ak}.

We also note that Airy functions and the quartic oscillator are connected through properties of the spectral determinant of the ($\al=0$) quartic oscillator \cite{Voros:1999rpg}, and one-dimensional chains have been related to the quartic ($\al=0$) oscillator via the well-known ODE/IM correspondence \cite{Dorey:2007zx}. The connection, if any, between the Fredholm operators presented in this paper and the results above about the spectral determinant remains to be explored.


\section{Acknowledgments}
The author is grateful to Carolina Figueiredo, Lawrence Hall, and Daniil Radamovich for discussions, and also grateful to Ariel Edery for helpful correspondence.
During the preparation of this work, \texttt{Mathematica} was used to assist with the algebra and numerical calculations, and Google's \texttt{Gemini} was used to assist with the algebraic derivation of Equations \eqref{eqn:muWKB}-\eqref{eqn:EWKB} and \eqref{eqn:ClausenF}, and for locating references and converting them to bibtex format. The author independently verified and validated all steps in the algebraic derivation for correctness, and the author also checked (and corrected where necessary) the validity of the references.
The Mathematica code that helped produce \tabref{tab:muECPsi} is available at \cite{ganor_2026_18969560}.




\begin{appendix}

\section{Derivation of the third moment}
\label{app:DerivationOfThirdMoment}

To derive \eqref{eqn:muMom3}, we begin by using the asymptotic expansion of the Airy function to derive the asymptotic expansion of $f(\aA)$ [defined in \eqref{eqn:faAdef}] as $\aA\rightarrow\infty$.
Using \texttt{Mathematica} we found
\bear
2\sqrt{2}e^{2\aA^{3/2}}f(\aA) &=&
\frac{1}{8 \aA^{3/2}}
-\frac{17}{128 \aA^3} 
+\frac{455}{2048 \aA^{9/2}} 
-\frac{50357}{98304 \aA^6} 
+\frac{2387359}{1572864 \aA^{15/2}} 
+\cdots
\label{eqn:asympfa}
\eear
It can be verified from the definition \eqref{eqn:faAdef}, with integration by parts, that $f(\aA)$ satisfies the ODE \eqref{eqn:7thOrderODE}.
If we look for solutions in the form
\be\label{eqn:fintgu}
f(\aA) = \int_{-\infty}^\infty e^{\frac{i}{3}u^3+i u x}g(u)du,
\ee
we find that $g(u)$ should satisfy
\be\label{eqn:ODEg}
18u g'''(u) + (16 i u^3 + 27) g''(u) + 48 i u^2 g'(u) + 16 i u g(u) = 0.
\ee
The general solution of \eqref{eqn:ODEg} is
\bear
g(u) &=& C_1 \cdot {}_2F_2\left( \frac{1}{3}, \frac{1}{3}; \frac{2}{3}, \frac{5}{6}; -\frac{8i u^3}{27} \right) + C_2 \cdot u^{1/2} \cdot {}_2F_2\left( \frac{1}{2}, \frac{1}{2}; \frac{5}{6}, \frac{7}{6}; -\frac{8i u^3}{27} \right) 
\nn\\ &&
+ C_3 \cdot u \cdot {}_2F_2\left( \frac{2}{3}, \frac{2}{3}; \frac{7}{6}, \frac{4}{3}; -\frac{8i u^3}{27} \right).
\label{eqn:gu}
\eear
We can obtain additional solutions by deforming the contour of integration in \eqref{eqn:fintgu} in the complex plane.
The asymptotic behavior of ${}_2F_2$ is \cite{NIST:HGPFQ}:
$$
{}_2F_2(a_1, a_2; b_1, b_2; z)
\sim
\left\{\begin{array}{ll}
e^z z^{a_1+a_2-b_1-b_2} & \text{for $\text{Re} z>0$} \\
z^{-\min_m a_m}  & \text{for $\text{Re} z<0$} \\
\end{array}\right.
$$
Using this expansion we can find rays in the complex plane along which the integrals of the pieces in \eqref{eqn:gu}, when plugged into \eqref{eqn:fintgu}, are convergent. Changing the integration variable $u$ to $t=e^{-i\phi}u$, for an appropriate constant phase $\phi$, we then find that the following integrals converge:
\bear
I_1 &=&
\int_0^\infty{}_2F_2\left( \tfrac{1}{3}, \tfrac{1}{3}; \tfrac{2}{3}, \tfrac{5}{6}; \tfrac{8}{27}t^3 \right) e^{-\frac{1}{3}t^3+i \aA e^{\frac{\pi i}{6}}t}dt
 \nn\\
I_2 &=&
\int_0^\infty{}_2F_2\left( \tfrac{1}{2}, \tfrac{1}{2}; \tfrac{5}{6}, \tfrac{7}{6}; \tfrac{8}{27}t^3 \right) 
t^{1/2}
 e^{-\frac{1}{3}t^3+i \aA e^{\frac{\pi i}{6}}t}dt
 \nn\\
I_3 &=&\int_0^\infty{}_2F_2\left( \tfrac{2}{3}, \tfrac{2}{3}; \tfrac{7}{6}, \tfrac{4}{3}; \tfrac{8}{27}t^3 \right)
t
 e^{-\frac{1}{3}t^3+i \aA e^{\frac{\pi i}{6}}t}dt
 \nn
\eear

\bear
I_4 &=&
\int_0^\infty{}_2F_2\left( \tfrac{1}{3}, \tfrac{1}{3}; \tfrac{2}{3}, \tfrac{5}{6}; \tfrac{8}{27}t^3 \right) e^{-\frac{1}{3}t^3+i \aA e^{\frac{5\pi i}{6}}t}dt
 \nn\\
I_5 &=&
\int_0^\infty{}_2F_2\left( \tfrac{1}{2}, \tfrac{1}{2}; \tfrac{5}{6}, \tfrac{7}{6}; \tfrac{8}{27}t^3 \right) 
t^{1/2}
 e^{-\frac{1}{3}t^3+i \aA e^{\frac{5\pi i}{6}}t}dt
 \nn\\
I_6 &=&\int_0^\infty{}_2F_2\left( \tfrac{2}{3}, \tfrac{2}{3}; \tfrac{7}{6}, \tfrac{4}{3}; \tfrac{8}{27}t^3 \right)
t
 e^{-\frac{1}{3}t^3+i \aA e^{\frac{5\pi i}{6}}t}dt
 \nn
\eear

\bear
I_7 &=&
\int_0^\infty{}_2F_2\left( \tfrac{1}{3}, \tfrac{1}{3}; \tfrac{2}{3}, \tfrac{5}{6}; \tfrac{8}{27}t^3 \right) e^{-\frac{1}{3}t^3+\aA t}dt
 \nn\\
I_8 &=&
\int_0^\infty{}_2F_2\left( \tfrac{1}{2}, \tfrac{1}{2}; \tfrac{5}{6}, \tfrac{7}{6}; \tfrac{8}{27}t^3 \right) 
t^{1/2}
 e^{-\frac{1}{3}t^3+\aA t}dt
 \nn\\
I_9 &=&\int_0^\infty{}_2F_2\left( \tfrac{2}{3}, \tfrac{2}{3}; \tfrac{7}{6}, \tfrac{4}{3}; \tfrac{8}{27}t^3 \right)
t
 e^{-\frac{1}{3}t^3+\aA t}dt,
 \nn
\eear
and
\be\label{eqn:SevenSols}
I_2, I_5, I_8, \quad
e^{\frac{\pi i}{6}}I_1 + e^{\frac{3\pi i}{2}}I_7,
\quad
e^{\frac{5\pi i}{6}}I_4 + e^{\frac{3\pi i}{2}}I_7,
\quad
e^{\frac{\pi i}{6}}I_3 + e^{\frac{3\pi i}{2}}I_9,
\quad
e^{\frac{5\pi i}{6}}I_6 + e^{\frac{3\pi i}{2}}I_9,
\ee
can be taken as a basis for the solutions of \eqref{eqn:7thOrderODE}. To find the solution we need, we expand for $\aA\rightarrow\infty$. For example, 
\bear
\lefteqn{
i\int_{-\infty}^\infty
u \cdot {}_2F_2\left(\tfrac{2}{3}, \tfrac{2}{3}; \tfrac{7}{6}, \tfrac{4}{3}; -\tfrac{8i u^3}{27} \right)
e^{\frac{i}{3}u^3+i u \aA}du
}\nn\\
&=&
\frac{\sqrt{\pi}\Gamma(\tfrac{7}{6})}{2^{5/3}\Gamma(\tfrac{2}{3})\Gamma(\tfrac{5}{6})\aA^{3/4}}\left(\sqrt{3}\pi-3\gamma-15\log 2-\tfrac{9}{2}\log\aA\right)e^{-\frac{2\aA^{3/2}}{3}}
+O\left(\frac{\log\aA}{\aA^{9/4}}\right)e^{-\frac{2\aA^{3/2}}{3}}.
\nn
\eear
Matching the linear combinations of these expressions with the asymptotic expansion \eqref{eqn:asympfa}, we arrive at the desired linear combination of solutions \eqref{eqn:SevenSols}, which reduces to \eqref{eqn:muMom3} when the integration contour is deformed back to the real axis.


\section{Explicit form of $\tr(\OpH\OpKMatH_0^{(\mr=2)})$}
\label{app:trHOpKMat}

To find an explicit expression for the integral \eqref{eqn:trHOpKMat2}, we first define
$$
\RGz_n(\zeta) = i^n\int_{-\infty}^\infty \frac{e^{\frac{i}{6}\wz^3+i\zeta\wz}d\wz}{\wz^n},
$$
which for $n=0,\dots,3$ evaluates to the following expressions.
$$
\RGz_0(\zeta)=
2^{4/3}\pi\Ai\left(2^{1/3}\zeta\right),
$$

\bear
\RGz_1(\zeta) &=&
-2^{1/3}\cdot 3^{-1/6}\Gamma\left(\tfrac{1}{3}\right)
{}_1F_2\left(\tfrac{1}{3}; \tfrac{2}{3},\tfrac{4}{3};\frac{2}{9}\zeta^3\right)\zeta
+2^{-1/3}\cdot 3^{1/6}\Gamma\left(\tfrac{2}{3}\right)
{}_1F_2\left(\tfrac{2}{3}; \tfrac{4}{3},\tfrac{5}{3};\frac{2}{9}\zeta^3\right)\zeta^2
+\frac{2\pi}{3},
\nn
\eear

\bear
\RGz_2(\zeta) &=&
2^{-1/3}\cdot 3^{1/6}\Gamma(\tfrac{2}{3})
-\tfrac{2}{3}\pi\zeta
+\tfrac{2}{3}\pi I_{-\frac{2}{3}}\left(\frac{1}{3}2^{3/2}\zeta^{3/2}\right)\zeta
\nn\\  &&
-2^{-2/3}3^{-1/6}\Gamma(\tfrac{1}{3}){}_1F_2(\tfrac{1}{3}; \tfrac{4}{3},\tfrac{5}{3}; \frac{2}{9}\zeta^3)\zeta^2
-2^{-1/3}3^{-1/6}\Gamma(\tfrac{2}{3}){}_1F_2(\tfrac{2}{3}; \tfrac{4}{3},\tfrac{5}{3}; \frac{2}{9}\zeta^3)\zeta^3,
\nn
\eear
and
\bear
\RGz_3(\zeta)
&=&
\tfrac{1}{3}\pi\zeta^2 
-\tfrac{1}{3}2^{1/2}\pi\sqrt{\zeta}I_{\frac{1}{3}}\left(\frac{1}{3}2^{3/2}\zeta^{3/2}\right)
+\tfrac{1}{3}2^{-1/2}\pi\sqrt{\zeta}I_{-\frac{1}{3}}\left(\frac{1}{3}2^{3/2}\zeta^{3/2}\right)
\nn\\ &&
-2^{-5/3}3^{-1/6}\Gamma(\tfrac{1}{3}){}_1F_2\left(\tfrac{1}{3}; \tfrac{4}{3}, \tfrac{5}{3}; \frac{2}{9}\zeta^3\right)\zeta^3
-2^{-7/3}3^{1/6}\Gamma(\tfrac{2}{3}){}_1F_2\left(\tfrac{2}{3}; \tfrac{5}{3}, \tfrac{7}{3}; \frac{2}{9}\zeta^3\right)\zeta^4.
\nn
\eear

Then, after integration, the expression \eqref{eqn:trHOpKMat2} takes the explicit form
\bear
\tr(\OpH\OpKMatH_0^{(\mr=2)})
&=&
\frac{2\bA\pi^3}{\evHK_1-\evHK_2}
\Bigl\lbrack
\muEVHK_2\RGz_0\left(\muEVHK_1\right)\RGz_2\left(\muEVHK_2\right)
+\RGz_2\left(\muEVHK_1\right)\RGz_1\left(\muEVHK_2\right)
+3\RGz_0\left(\muEVHK_1\right)\RGz_3\left(\muEVHK_2\right)
\nn\\ &&
-\RGz_1\left(\muEVHK_1\right)\RGz_2\left(\muEVHK_2\right)
-\muEVHK_1\RGz_2\left(\muEVHK_1\right)\RGz_0\left(\muEVHK_2\right)
-3\RGz_3\left(\muEVHK_1\right)\RGz_0\left(\muEVHK_2\right)
+(\muEVHK_1-\muEVHK_2)\RGz_1\left(\muEVHK_1\right)\RGz_1\left(\muEVHK_2\right)
\Bigr\rbrack,
\nn
\eear
where we defined
$$
\muEVHK_k\equiv\frac{\evHK_k}{4\bA^2}\,.
$$

\end{appendix}

\bibliographystyle{apsrev4-2}
\bibliography{refs} 

\end{document}